\documentclass[aps,prl,twocolumn,showpacs,superscriptaddress]{revtex4-1}
\usepackage{graphicx}
\usepackage{dcolumn}
\usepackage{bm}
\usepackage{amssymb}
\usepackage{pifont}
\usepackage{bbding}
\usepackage[usenames,dvipsnames]{color}
\usepackage{placeins}
\usepackage{hyperref}
\usepackage{ulem}

\definecolor{grin}{RGB}{19,163,19}

\newcommand{\tp}{t}

\begin{document}

 \normalem

\title{Suppression and Revival of Weak Localization through Control of Time-Reversal Symmetry}

\author{K. M\"uller}
\author{J. Richard}
\author{V. V. Volchkov}
\author{V. Denechaud}
\affiliation{Laboratoire Charles Fabry UMR 8501,
Institut d'Optique, CNRS, Univ Paris Sud 11,
2 Avenue Augustin Fresnel,
91127 Palaiseau cedex, France}
\author{P. Bouyer}
\affiliation{LP2N UMR 5298,
Univ Bordeaux 1, Institut d'Optique and CNRS,
351 cours de la Lib\'eration,
33405 Talence, France.}
\author{A. Aspect}
\author{V. Josse}
\email{vincent.josse@institutoptique.fr}
\affiliation{Laboratoire Charles Fabry UMR 8501,
Institut d'Optique, CNRS, Univ Paris Sud 11,
2 Avenue Augustin Fresnel,
91127 Palaiseau cedex, France}

\begin{abstract}
We report on the observation of suppression and revival of coherent backscattering of ultra-cold atoms launched in an optical disorder in a quasi-2D geometry and submitted to a short dephasing pulse, as proposed in T. Micklitz \textit{et al.}, Phys. Rev. B \textbf{91}, 064203 (2015).
This observation demonstrates a novel and general method to study weak localization by manipulating time reversal symmetry in disordered systems. 
In future experiments, this scheme could be extended to investigate higher order localization processes at the heart of Anderson (strong) localization.
\end{abstract}


\maketitle

\begin{figure}[b]
\includegraphics[width=0.45\textwidth]{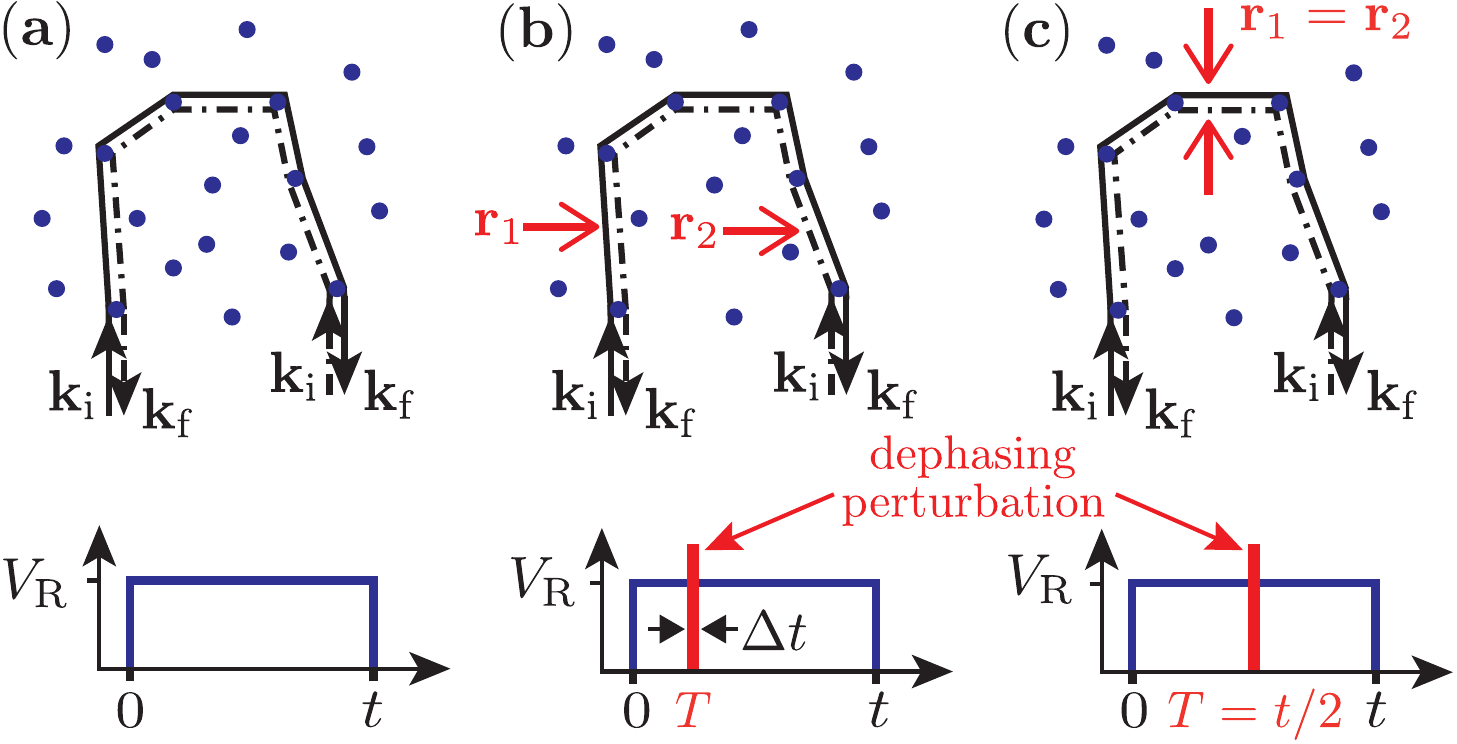}
\caption{
\label{fig:CBSE_Explanation}
\textbf{Principle of CBS suppression and revival following a dephasing kick.}
\textbf{(a)}~CBS results from the constructive interference between the scattering amplitudes of direct (1, solid line) and reciprocal (2, dashed) multiple scattering paths, for a plane matter wave launched with a wave vector $\mathbf{k}_\mathrm{i}$ and detected in the backward direction, at $\mathbf{k}_\mathrm{f}=-\mathbf{k}_\mathrm{i}$.
The time diagram shows the switching on at~$0$, and off after time~$t$ of the disordered potential (amplitude $V_\mathrm{R}$). 
\textbf{(b)}~CBS suppression: A pulsed perturbation at time $T\neq \tp/2$ entails a phase difference [$\phi_\mathrm{kick}(\mathbf{r}_1)\neq \phi_\mathrm{kick}(\mathbf{r}_2)$] between the amplitudes associated with paths 1 and 2, where $\mathbf{r}_1$ and $\mathbf{r}_2$ are the positions on each path at time $T$.
The constructive interference is destroyed.
\textbf{(c)}~CBS revival: For the special case $T=\tp/2$, one has $\mathbf{r}_1=\mathbf{r}_2$ and the constructive interference between the direct and reciprocal paths is restored.
A CBS revival is then expected at $t=2T$.}
\end{figure}

Weak localization is a fundamental phenomenon in mesoscopic physics.
It arises from constructive interferences between \textit{time-reversed} multiple scattering paths that modify the transport properties of (matter) waves in disordered media.
For electronic systems, these interferences increase the probability that an electron remains around its initial position, thereby acting against propagation and resulting in an increased resistivity~\cite{Abrahams1979,Gorkov1979,Bergmann1984weak}.
In AMO physics, these interferences lead to the coherent backscattering (CBS) phenomenon~\cite{Tsang1984,Akkermans1986}, that is an enhancement of the scattering probability in the backward direction when a plane wave is launched into a disordered medium [see Fig.~\ref{fig:CBSE_Explanation}\textbf{(a)}].
This was first observed with light~\cite{Kuga1984,Albada1985,Wolf1985}, and then with a large variety of classical waves (see e.g.~\cite{Akkermans2007} and references therein).

Weak localization is extremely sensitive to any perturbation of the time reversal symmetry of the wave propagation~\footnote{In principle, reciprocity is sufficient to ensure weak localization~\cite{Tiggelen1998}.
In absence of dissipation, as it is the case in the experiment presented here, reciprocity and time reversal symmetry are however equivalent~\cite{Carminati2000}.}.
For electrons, it is therefore suppressed in the presence of spin-orbit coupling or magnetic impurities~\cite{Hikami1980}, or when a time dependent potential is applied to the system (e.g., using high frequency rf fields)~\cite{altshuller1981suppression,*wei2006microwave}.
In turn, the controlled manipulation of time reversal symmetry provides direct signatures of phase coherence in condensed matter physics. 
Electrons are sensitive to external magnetic fields, which induce, via the accumulated Aharonov-Bohm phase, a dephasing between the counter-propagating paths responsible for weak localization.
Depending on the geometry, this leads either to the magneto-negative resistance effect in thin metallic films~\cite{Bergmann1984weak}, or to the oscillations of the resistivity in thin walled cylinders~\cite{Altshuler1981,*sharvin1981magnetic,*al1982observation}.
In direct analogy, the suppression of CBS under time reversal symmetry breaking has been studied with classical waves, as for instance with acoustic waves in a rotating medium~\cite{Derosny2005}, with light in a Faraday medium~\cite{Erbacher1993,Lenke2000} or recently in a disordered medium subjected to an ultrafast modulation~\cite{Muskens2012}.

In this letter, we use neutral ultracold atoms to demonstrate a novel method, as proposed by T.~Micklitz and coworkers~\cite{Micklitz2014}, to manipulate time reversal symmetry and study weak localization. In our system, the atoms propagate in an optical disordered potential in a quasi-2D configuration. CBS manifests itself as a backward peak in the momentum distribution of the scattered atoms~\cite{Cherroret2012,Jendrzejewski2012,Labeyrie2012}.
Time reversal symmetry is broken by applying, at a chosen time $T$, a quasi-instantaneous dephasing kick on the atoms.
Following this kick, we observe the suppression of the CBS peak, except for a revival that happens at time $2T$.
This revival constitutes a new direct signature of phase coherence in disordered systems.

\begin{figure}[b]
\includegraphics[width=0.45\textwidth]{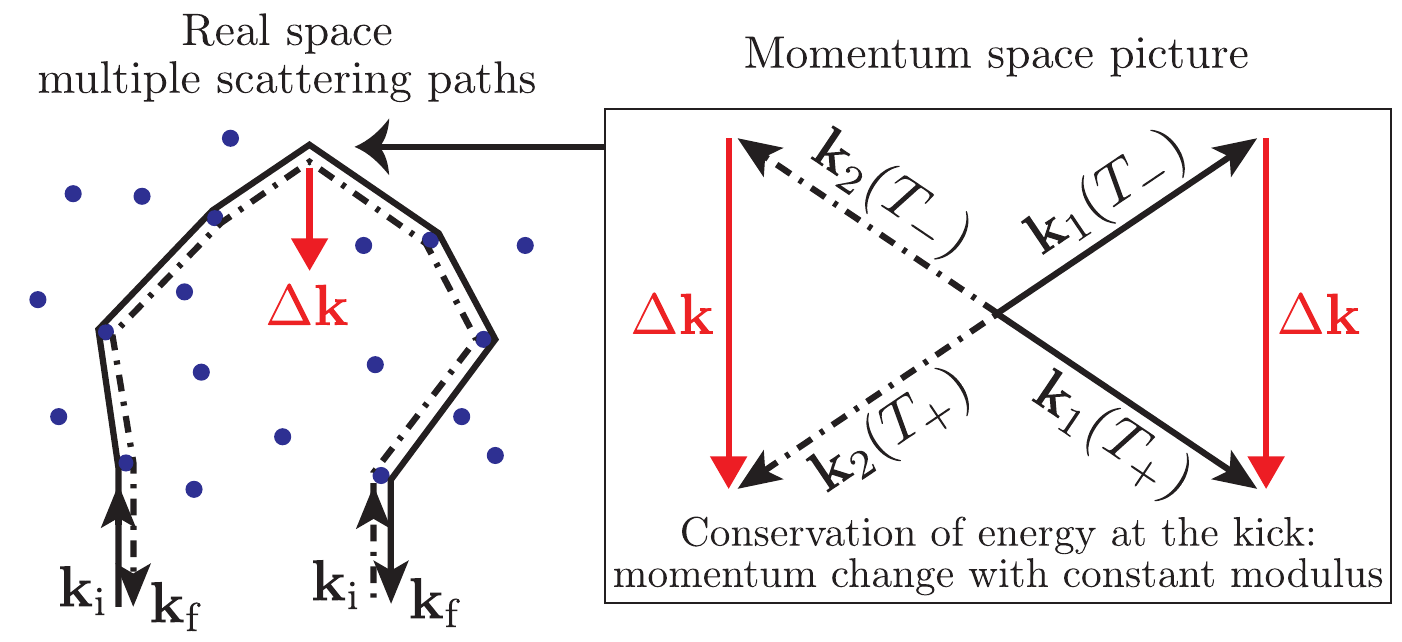}
\caption{\label{fig:CBSE_Explanation2}
\textbf{Generalization of the revival mechanism for strong momentum changes.}
All multiple scattering paths that participate to the momentum distribution at $\mathbf{k}_\mathrm{f}=-\mathbf{k}_\mathrm{i}$ must conserve the modulus of the momentum at the kick:
$\vert \mathbf{k}(T_+)\vert = \vert \mathbf{k}(T_-)+\Delta \mathbf{k} \vert =\vert \mathbf{k}(T_-)\vert$, where $T_-$ and $T_+$ are the times just before and after the kick (see inset).
For the observation time $t=2T$, each direct path (solid line) satisfying that condition has a time-reversed counterpart (dashed line) satisfying the same condition, and for which the momentum change happens at the same position.
The constructive interference between these reciprocal paths leads to the CBS revival at $\mathbf{k}_\mathrm{f}=-\mathbf{k}_\mathrm{i}$, ideally with a contrast of unity.}
\end{figure}

The principle of the suppression and revival effect is illustrated in Fig.~\ref{fig:CBSE_Explanation}, where we consider atoms with initial momentum $\hbar \mathbf{k}_\mathrm{i}$, launched into a disordered potential at time $t=0$, and whose momentum distribution is observed at time $t$.
In the absence of any perturbation, the constructive interferences between counter-propagating multiple scattering paths lead to an enhancement of the scattering probability towards $\mathbf{k_\mathrm{f}}=-\mathbf{k_\mathrm{i}}$ [Fig.~\ref{fig:CBSE_Explanation}(a)].
That is the CBS peak.
At a chosen time $T$ during the propagation in the disorder ($0<T<\tp$), we apply an inhomogeneous potential $V(\mathbf{r})$ on the atoms [Fig.~\ref{fig:CBSE_Explanation}(b)], for a duration $\Delta t$ short enough that the perturbation can be considered as an instantaneous kick.
It imprints, on each multiple scattering amplitude, a \textit{local} phase $\phi_\mathrm{kick} (\mathbf{r})=V(\mathbf{r}) \Delta t/\hbar$ that depends on the position of the atom $\mathbf{r}$ at the time $T$~\footnote{We use a WKB-like approximation to calculate the phase of the scattering amplitude along a scattering path}.
It also entails a momentum change $\hbar \Delta \mathbf{k}= - \mathbf{\nabla} V(\mathbf{r}) \Delta t$.
If this change is small, one can neglect the subsequent modifications of the atomic trajectories.
The overall effect of the kick is then to introduce a controlled dephasing between the counter-propagating paths~\footnote{We consider in Eq.~(\ref{Eq:dephasing}) the case of a potential gradient uniform at the scale of the atomic cloud, as it is the case in the experiment.},
\begin{equation}
\Delta \phi_\mathrm{kick}= \phi_\mathrm{kick}(\mathbf{r}_2)-\phi_\mathrm{kick} (\mathbf{r}_1)= \mathbf{\Delta k}\cdot (\mathbf{r}_2-\mathbf{r}_1),
\label{Eq:dephasing}
\end{equation}
where $\Delta$k will be referred to as the strength or the amplitude of the kick.
$\mathbf{r}_{(1,2)}$ correspond to the respective positions of the atoms on each path at time $T$.
In general, these two positions are separated in space, and the dephasings $\Delta \phi_\mathrm{kick}$ are finite.
The averaging of these dephasings over all possible multiple scattering paths leads to the suppression of the CBS peak.

If, however, the kick is applied at $T=\tp/2$, the two positions $\mathbf{r}_{1}$ and $\mathbf{r}_{2}$ always coincide [Fig.~\ref{fig:CBSE_Explanation}(c)], so that the dephasing in Eq.~(\ref{Eq:dephasing}) is null and the constructive interference between the counter-propagating paths is restored.
Altogether, we therefore expect a suppression of the CBS peak after the kick, except for a revival at time $\tp=2T$~\cite{Micklitz2014}.
In general terms, this revival can be traced back to the brief re-establishment of the time reversal symmetry around the time $2T$.

\begin{figure}[t]
\includegraphics[width=0.45\textwidth]{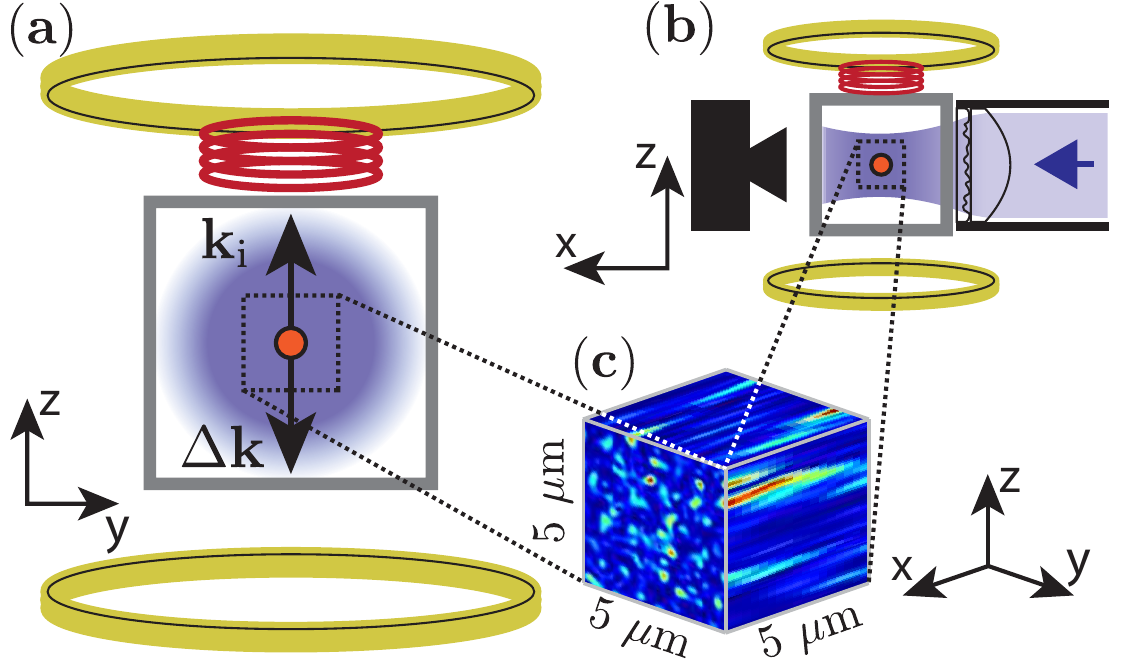}
\caption{\label{fig:ExpSetup} \textbf{Experimental Setup.}
\textbf{(a)} and \textbf{(b)}: A cloud of ultracold atoms (orange disk) is launched upwards, along the $z$ axis, with an initial momentum~$\hbar\mathbf{k}_\mathrm{i}$.
The disordered potential is created by an anisotropic laser speckle, elongated along the $x$ axis, obtained by passing a laser beam through a scattering plate (shaded blue).
The atoms are suspended against gravity by magnetic levitation provided by the pair of large coils (yellow).
The small red coil is used to create a magnetic gradient pulse resulting in a downward momentum change $\hbar\Delta \bold{k}$, along the $z$ axis, to the atoms. After a time of flight of 150~ms, fluorescence imaging recorded by an EMCCD camera along $x$ gives the transverse momentum distribution (in the $y-z$ plane, see text).
\textbf{(c)} 3D false color representation of the disordered potential.}
\end{figure}

\begin{figure*}
\includegraphics[width=1\textwidth]{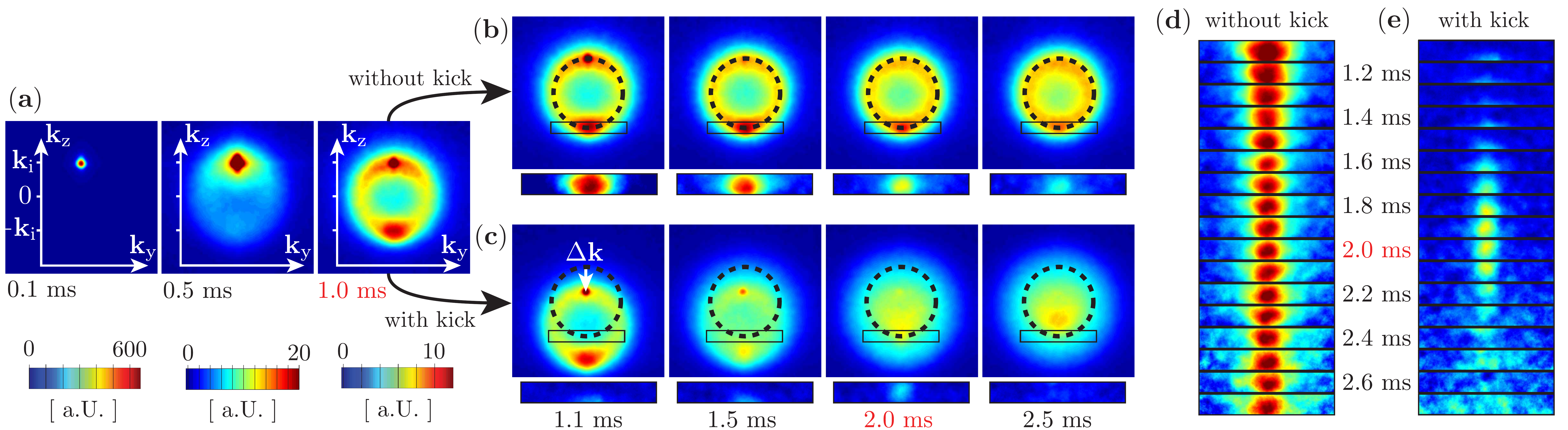}
\caption{\label{fig:CoherentPartEvolution}\textbf{Observation of the CBS revival}.
The images represent the 2D \textbf k-vector distribution in the $y-z$ plane, $n(\bold{k}, t)$.
Each image results from an average over 30 experimental runs and the color scale is kept unchanged, except for the first two images in \textbf{(a)}.
\textbf{(a)} Initial evolution of $n(\bold{k}, t)$ for atoms launched at $t=0$ in the disorder with momentum $\hbar \mathbf{k}_\mathrm i$.
The initial peak at $\mathbf{k}_\mathrm i$ decays while the atoms' {\bf k}-vectors are redistributed on a ring of radius $|k_\mathrm{i}|$, and the CBS peak grows at $-\mathbf{k}_\mathrm{i}$.
\textbf{(b)} and \textbf{(c)}: Evolution for $t>T$, without and with the dephasing kick at $T=1$~ms.
The dashed circles are centered on the origin, and have a radius $|k_\mathrm{i}|$.
The amplitude of the kick $\Delta k$ is measured from the first image in \textbf{(c)}.
The rectangular boxes below each image show the extracted coherent fraction $\mathcal{C}_\mathrm{coh}(\mathbf{k},t)$ around $-\mathbf{k}_\mathrm{i}$ (see text).
\textbf{(d)} and \textbf{(e)} Coherent fractions around $-\mathbf{k}_\mathrm{i}$, as in \textbf{(b)} and \textbf{(c)}, but renormalized by the reference CBS peak at the same time (see text).}
\end{figure*}

Let us note that the discussion above, valid for a small momentum change as considered in~\cite{Micklitz2014}, can be generalized to the case of a strong momentum change, where the modifications of the atomic trajectories at the dephasing kick have to be taken into account (see Fig.~\ref{fig:CBSE_Explanation2}).
A CBS revival, at $\mathbf{k}_\mathrm{f}=-\mathbf{k}_\mathrm{i}$  and time $t=2T$, is then expected whatever the kick's strength.

The experimental set-up, sketched in Fig.~\ref{fig:ExpSetup}, has been described in~\cite{Jendrzejewski2012}.
A salient feature is the ability to launch into the disorder a ``quasi-monochromatic'' cloud of non-interacting neutral atoms, i.e., an ensemble of atoms with a narrow velocity spread.
The cloud, containing $10^5$ $^{87}$Rb atoms is prepared in the paramagnetic hyperfine Zeeman sub-level $|F=2,m_F=-2\rangle$, allowing us to exactly compensate gravity \footnote{We use the magnetic dipole force, created by a vertical magnetic field gradient, to compensate gravity.}.
The atoms are launched along $z$ with a mean velocity of $v_\mathrm{i}=3.09\pm 0.04~\text{mm/s}$ ($k_\mathrm{i} = 4.24~\mu\text{m}^{-1}$, where $\mathbf{k}=m \mathbf v/\hbar$ and $m$ is the atom mass), and a velocity spread of $0.18\pm 0.02~\text{mm/s}$.
At $t=0$, we switch on an anisotropic speckle field, elongated along the $x$ axis, created by a far off resonance laser (wavelength 532~nm)~\cite{Clement2006,Goodman2007}.
The resulting disordered potential has correlation lengths much shorter in the $y$-$z$~plane than in the $x$ direction ($\sigma_y=\sigma_z=0.27~\mu\text{m}$ and $\sigma_x=1.40~\mu\text{m}$ HWHM), so that the situation can be considered 2D for atoms launched in the $y$-$z$~plane~\cite{Jendrzejewski2012}.
At time $t$, the disorder is switched off and the 2D atomic velocity distribution [or equivalently the $\mathbf{k}$-vector distribution $n(\mathbf{k},t)$ with $\mathbf{k}$ in the $y$-$z$~plane] is recorded by fluorescence imaging, along the $x$ axis, after a long free expansion time of 150~ms permitted by the gravity compensation.
Taking into account the initial size of the cloud, the overall resolution is estimated to be $\Delta k_\mathrm{res}=0.3$~$\mu$m$^{-1}$.

The dephasing kick described above is created by a pulsed magnetic gradient $B'_\mathrm{kick}(t)$ along the $z$ axis (Fig.~\ref{fig:ExpSetup}), for a typical duration $\Delta t\sim35~\mu$s.
It imposes an inhomogeneous potential $V(z)=-\mu_\mathrm{B}B'_\mathrm{kick} z$ where $B'_\mathrm{kick}\sim -100$~G/cm.
The precise amplitude $\Delta k$ of the kick is directly measured in the experiment (see below).

Figure~\ref{fig:CoherentPartEvolution} shows the evolution of the \textbf{k}-vector distribution when atoms are subjected to the disordered potential, with or without the dephasing kick applied.
The average disorder amplitude is set to $V_\mathrm{R}/h= 660$~Hz~\footnote{For laser speckles the average value of the disorder coincides with its standard deviation~\cite{Goodman2007}.}.
Panels \textbf{(a)} and \textbf{(b)} correspond to the reference case, i.e., when no kick is applied ($\Delta \bold{k}=0$).
In the beginning, the initial peak centered at $\mathbf{k}_\mathrm{i}$ decays, while the momenta are redistributed on a ring of mean radius $|k_\mathrm{i}| $ (elastic scattering).
Monitoring that decay and the isotropization of the momentum distribution, we infer the mean scattering time $\tau_\mathrm{s}= 0.22~\text{ms}$ and the transport time $\tau^\star = 0.6~\text{ms}$~\cite{Plisson2013}.
After a few scattering events the CBS peak develops around $\mathbf{k}_\mathrm{f}=-\mathbf{k}_\mathrm{i}$ and becomes clearly visible at $t=1$~ms.
As explained in~\cite{Jendrzejewski2012}, we observe a decrease of the contrast of the CBS peak for longer times, which can be attributed to two reasons.
Firstly, the CBS peak width becomes smaller than the experimental resolution.
Secondly, after a few scattering events the probability for an atom to scatter out of the $y$-$z$ plane becomes significant.
As a consequence, the observation time of the CBS peak dynamics is limited to about $6~\tau^\star$ (3.5~ms) in our experiment.

Panel \textbf{(c)} of Fig.~\ref{fig:CoherentPartEvolution} shows the evolution of the momentum distribution after the dephasing kick, with the connected momentum change $\hbar \mathbf{ \Delta k}$ along $z$, is applied.
The whole \textbf{k}-vector distribution, including the CBS peak, is suddenly shifted downwards (first image) by $\Delta k =- 3.44 \pm 0.3$~$\mu$m$^{-1}$ ($\Delta k \sim - 0.8 \,k_\mathrm{i}$).
Then the momenta are redistributed by elastic scattering, and, after a large enough number of scattering events, the momentum distribution tends towards a broad, isotropic distribution.
During that process, the CBS peak is rapidly suppressed~\footnote{A large value of the kick amplitude $\Delta k$ is chosen to ensure a rapid dephasing.
After one scattering time $\tau_\mathrm{s}$, the dephasing can be as large as $\Delta \phi_\mathrm{kick}\sim \Delta k \, v \tau_\mathrm{s}= 2.3>1$ [see Eq.~(\ref{Eq:dephasing})].
During that dephasing, the position of the coherent peak has a dynamic in momentum space~\cite{Micklitz2014}.
This dynamic, visible on Fig.~\ref{fig:CoherentPartEvolution}, will be analyzed in a future work.}.

As discussed above, a revival of the CBS peak is expected to appear at $t=2T=2$~ms around $\mathbf k_\mathrm{f}=-\mathbf k_\mathrm{i}$, on top of an incoherent background.
In order to reveal it, we first estimate the incoherent background $n_\mathrm{incoh}(\bold{k},t)$ by performing a quadratic fit of the distribution outside a rectangular box centered on $-\mathbf k_\mathrm i$ and further extrapolating it into that box.
The coherent fraction, defined as $\mathcal{C}_\mathrm{coh} (\mathbf{k},t) = [n(\mathbf{k},t)-n_\mathrm{incoh}(\mathbf{k},t)] / n_\mathrm{incoh}(\mathbf{k},t)$, is then extracted.
Panel \textbf{(e)} shows that coherent fraction after application of the kick, hereafter referred to as $\mathcal{C}_\mathrm{kick} (\mathbf{k},t)$ [for comparison, the coherent fraction in the absence of the kick, $\mathcal{C}_\mathrm{ref}(\mathbf{k},t)$, is shown in \textbf{(d)}, and in both panels the coherent fractions are normalized by $\mathcal{C}_\mathrm{ref}(-\mathbf{k}_\mathrm{i},t)$]. 
A clear revival of the CBS peak is observed around $t=2$~ms, a striking evidence of the predicted phenomenon.

\begin{figure}[t]
\includegraphics[width=0.45\textwidth]{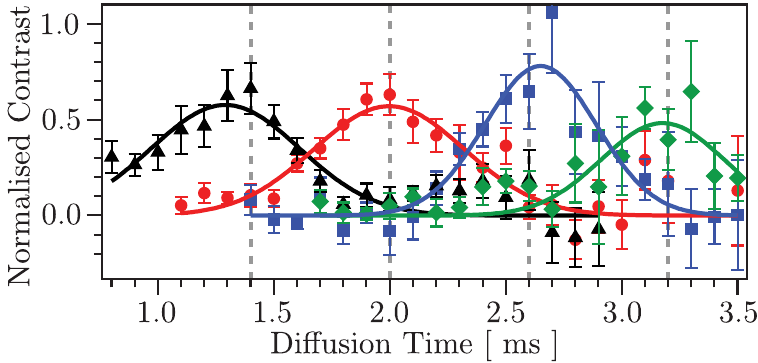}
\caption{\label{fig:OnShellContrast}\textbf{Amplitude of the CBS revivals $\gamma_\mathrm{rev}(t)$ for various choices of the dephasing times $T$}.
The data points \TriangleUp \textcolor{red}{\CircleSolid} \textcolor{blue}{\SquareSolid} \textcolor{OliveGreen}{\DiamondSolid} correspond to $T=$~0.7, 1.0, 1.3, and 1.6~ms respectively, the kick's strength $\Delta k$ being fixed (same as in Fig.~\ref{fig:CoherentPartEvolution}). 
The dotted vertical lines indicate the times $2 T$ when the revivals are expected.
Solid lines are Gaussian fits with zero offset.
The observed revival times, determined by the fits, are respectively (in ms): 1.3 $\pm$ 0.08, 1.99 $\pm$ 0.08, 2.65$\pm$ 0.05 and 3.18$\pm$ 0.09, in good agreement with the predicted values.
Uncertainties correspond to the 95\% confidence intervals.}
\end{figure}

Figure~\ref{fig:OnShellContrast} shows the temporal evolution of the revival, whose amplitude is defined as $\gamma_\mathrm{rev}(t)=\mathcal{C}_\mathrm{kick}(\bold{-k_\mathrm{i}}, t)/\mathcal{C}_\mathrm{ref}(\bold{-k_\mathrm{i}}, t)$, for several values of the kick time $T$ and the same kick strength $\Delta k$.
In all cases, a revival is observed around the expected $2T$ time, with a contrast of about 60$\%$.
We suspect that spurious magnetic fields, due to eddy currents that are excited by the magnetic pulse used to kick the atoms, are responsible for the reduced contrast of the revival. 
The revival shape is well fitted by a Gaussian with almost the same \emph{rms} width for all cases, about $\Delta \tau_\mathrm{rev}=0.28\pm 0.04~\text{ms}$.

We render an account of the shape and the width of the revival peak by considering the phase difference $\Delta \phi_\mathrm{kick}$ between the counter-propagating scatterings paths given by Eq.~(\ref{Eq:dephasing}).
The amplitude of the revival at $\mathbf{k}_f=-\mathbf{k}_i$ is expected to vary as $\gamma(t)=\langle \exp [\mathrm i \Delta \phi_\mathrm{kick} (\mathbf{R})]\rangle $, where the brackets indicate a statistical average over the separation $\mathbf{R}(t)=\mathbf r_1 -\mathbf r_2$, for $T$ fixed.
This separation is null at $t=2T$, so that $\gamma(t)=1$, i.e. the revival peak is maximal.
For $t\neq 2T$, $\mathbf{R}(t)$ is the distance corresponding to the propagation of the atoms in the disorder for a duration $|t-2T|$.

If this propagation was a fully developed random walk, the distance $\mathbf{R(t)}$ would have a Gaussian distribution, with variance $\langle \mathbf{R}^2 \rangle= 2D |t-2T|$, where $D=v_\mathrm{i}^2 \tau^\star/2$ is the diffusion constant.
The amplitude of the revival would then be
\begin{equation} 
\gamma_{\mathrm{dif}}(t)=e^{-\Delta \mathbf{k}^2 \langle \mathbf{R}(t)^2\rangle/2 }=e^{-\Delta \mathbf k^2 D \vert t- 2T\vert} .
\label{Eq:contrastdif}
\end{equation}
This formula is identical to the one derived by Micklitz \textit{et al.}~\cite{Micklitz2014}.
It predicts a profile with a symmetric exponential shape, of half-width (at $1/e$) $\Delta \tau_\mathrm{dif}= [D \Delta \mathbf k^2]^{-1} \sim 0.03$~ms for our parameters.
It does not correspond to our observation of a Gaussian shape (see Fig.~\ref{fig:OnShellContrast}), with a width one order of magnitude larger.

Actually, in our experiment, the hypothesis of a fully developed random walk is not fulfilled at the short time scale characterizing the revival width, which is on the order of a few scattering time $\tau_\mathrm{s}$.
A similar failure of the diffusive hypothesis was already observed in~\cite{Jendrzejewski2012} and we had rendered an excellent account of the observation at short times by using an effective ballistic dynamics, derived from \cite{Gorodnichev1994}, in which $\langle \mathbf{R}^2 \rangle= (v_\mathrm{i} \vert t-2T\vert/3 )^2$.
Using the same ansatz here, we obtain a Gaussian expression for the CBS revival profile:
\begin{equation}
\label{eq:BallRevival}
\gamma_\mathrm{bal}(t)= e^{-(t - 2T)^2/2\Delta \tau_\mathrm{bal}^2} ~\text{, where}~~\Delta\tau_\mathrm{bal} = \frac{3}{|\Delta k | v_\mathrm{i}} .
\end{equation}
For our parameters, one has $\Delta \tau_\mathrm{bal} = 0.28~\text{ms}$, in striking agreement with the observations.
In a series of supplementary measurements, we have varied the strength of the kick $\Delta k$, and the revival widths were found in good agreement with expression~(\ref{eq:BallRevival}) \footnote{To be published elsewhere.}.

In conclusion, we have demonstrated experimentally a new method to break and restore time reversal symmetry in a disordered medium, resulting in the suppression and revival of the CBS peak.
This observation provides an indisputable signature of weak localization and could be observed with any kind of waves.
For cold atoms, it could for instance serve to differentiate CBS from classical echoes as reported in~\cite{Labeyrie2012}. 
It could also be implemented differently, for instance by kicking the disordered potential itself, or with any kind of inhomogeneous external potential.
It would be interesting to compare it to schemes using time dependent potentials~\cite{DErrico2013}, or artificial gauge fields developed for neutral atoms~\cite{Towers2013}.
Finally, extending the scheme to multiple kick sequences opens new prospects to study Anderson (strong) localization in a renewed perspective~\cite{Micklitz2014}.
Depending on the chosen sequence, suppression and revival of both the CBS and of the expected coherent forward scattering peak~\cite{Karpiuk2012,*Micklitz2013,*Lee2014,*Ghosh2014} could be observed and used to investigate higher-order mechanisms at the heart of Anderson localization.
Such observations would ideally complement previous studies of Anderson localization with ultracold atoms~\cite{Billy2008,*Roati2008,Kondov2011,*Jendrzejewski2012b,*Semeghini2014}.

\begin{acknowledgments}
We thank T. Micklitz, C. M\"uller and A. Altland for their private communication of their proposal of~\cite{Micklitz2014}. Our research was supported by
 ERC (Advanced Grant ``Quantatop"),
 ANR (DisorderTransitions),
 DGA, R\'egion Ile de France (IFRAF) and Institut Universitaire de France.
\end{acknowledgments}


%

\end{document}